\documentclass[conference]{IEEEtran}
\IEEEoverridecommandlockouts
\usepackage{cite}
\usepackage{amsmath,amssymb,amsfonts}
\usepackage{algorithmic}
\usepackage{graphicx}
\usepackage{textcomp}
\usepackage{xcolor}
\def\BibTeX{{\rm B\kern-.05em{\sc i\kern-.025em b}\kern-.08em
    T\kern-.1667em\lower.7ex\hbox{E}\kern-.125emX}}

\makeatletter
\newcommand{\linebreakand}{%
  \end{@IEEEauthorhalign}
  \hfill\mbox{}\par
  \mbox{}\hfill\begin{@IEEEauthorhalign}
}
\makeatother

\usepackage{subcaption}
\usepackage{paralist}
\usepackage{pifont}
\usepackage{algorithm}
\usepackage{xspace}
\usepackage{caption}
\usepackage{url}
\usepackage{hyperref}
\usepackage{tikz}
\usepackage{thmbox}
\usepackage{multirow}

\newcommand{\name}{{\textbf{Hermes}}\xspace}
\newcommand{\mechanism}{{\textsc{PipeLoad}}\xspace}
\newcommand{\IA}{{\textsl{Inference Agent}}\xspace}
\newcommand{\DA}{{\textsl{Daemon Agent}}\xspace}
\newcommand{\LAs}{{\textsl{Loading Agents}}\xspace}
\newcommand{\LA}{{\textsl{Loading Agent}}\xspace}
\newcommand{\LP}{{\textbf{\textit{Layer Profiler}}}\xspace}
\newcommand{\PP}{{\textbf{\textit{Pipeline Planner}}}\xspace}
\newcommand{\EE}{{\textbf{\textit{Execution Engine}}}\xspace}

\newcommand{\circled}[1]{%
  \tikz[baseline=(char.base)]{
    \node[shape=circle,fill=black, minimum size=18pt, font=\Large,scale=0.5,text=white,draw,inner sep=-0.5pt] (char) {#1};
  }%
}

\begin{document}

\title{\name: Memory-Efficient Pipeline Inference for Large Models on Edge Devices\\
\thanks{This work was supported by Shanghai Key Laboratory of Scalable Computing and Systems, National Key Laboratory of Ship Structural Safety, and the Eighth Research Institute of China Aerospace Science and Technology Group Company, Ltd., under Grant USCAST2023-17 and Grant USCAST2023-21. (Corresponding author: Ruhui Ma.)}
}

\author{
    \IEEEauthorblockN{Xueyuan Han$^{1}$, Zinuo Cai$^{2}$, Yichu Zhang$^{3}$, Chongxin Fan$^4$, Junhan Liu$^{2}$, Ruhui Ma$^{2}$, Rajkumar Buyya$^{5}$}
    \IEEEauthorblockA{$^1$ ParisTech Elite Institute of Technology, Shanghai Jiao Tong University, Shanghai, China}
    \IEEEauthorblockA{$^2$ School of Electronic Information and Electrical Engineering, Shanghai Jiao Tong University, Shanghai, China}
    \IEEEauthorblockA{$^3$ UM-SJTU Joint Institute, Shanghai Jiao Tong University, Shanghai, China}
    \IEEEauthorblockA{$^4$ Shanghai Aerospace System Engineering Institute, Shanghai, China}
    \IEEEauthorblockA{$^5$ Cloud Computing and Distributed Systems (CLOUDS) Laboratory, The University of Melbourne, Melbourne, Australia}
    \IEEEauthorblockA{\{hxy771126, kingczn1314, 1654468697\}@sjtu.edu.cn, fcs-841220@163.com, \\ \{liujunhan, ruhuima\}@sjtu.edu.cn, rbuyya@unimelb.edu.au}
}


\maketitle

\begin{abstract}
The application of Transformer-based large models has achieved numerous success in recent years. 
However, the exponential growth in the parameters of large models introduces formidable memory challenge for edge deployment. 
Prior works to address this challenge mainly focus on optimizing the model structure and adopting memory swapping methods. However, the former reduces the inference accuracy, and the latter raises the inference latency. 
This paper introduces \mechanism, a novel memory-efficient pipeline execution mechanism. It reduces memory usage by incorporating dynamic memory management and minimizes inference latency by employing parallel model loading. Based on \mechanism mechanism, we present \name, a framework optimized for large model inference on edge devices. We evaluate \name on Transformer-based models of different sizes. Our experiments illustrate that \name achieves up to \(4.24\times\) increase in inference speed and \(86.7\%\) lower memory consumption than the state-of-the-art pipeline mechanism for BERT and ViT models, \(2.58\times\) increase in inference speed and \(90.3\%\) lower memory consumption for GPT-style models.
\end{abstract}

\begin{IEEEkeywords}
Edge computing, Memory optimisation, Large model inference, Pipeline execution.
\end{IEEEkeywords}
\section{Introduction}

The Transformer architecture has profoundly transformed the landscape of deep learning and brought forward large models with their applications spreading from data centers~\cite{cai2023smss} to edge devices.
Large models are generally categorized into Natural Language Processing (NLP), Computer Vision (CV), and Multimodal models.
NLP is widely applied on mobile devices~\cite{chen2020deep,murshed2021machine}, from intelligent personal assistants, like Google Assistant and Apple Siri to real-time language translation~\cite{ren2020simulspeech}.
CV plays a pivotal role in the field of autonomous driving~\cite{muhammad2020deep,deng2021deep}, where it is utilized for tasks such as real-time object detection~\cite{wang2023yolov7,cai2021yolobile}, lane recognition~\cite{tang2021review}, and traffic signal detection~\cite{ayachi2020traffic}.
By enriching robots' perception and decision-making capabilities through the integration of diverse data types~\cite{lindqvist2022multimodality}, such as visual, auditory~\cite{ince2021audiovisual}, and tactile~\cite{cao2024multimodal} information, Multimodal large models are revolutionizing the field of robotics~\cite{wu2023multimodal,xu2023multimodal}.



Due to the explosive growth in the size of large models, deploying them at the edge faces critical memory challenges~\cite{alizadeh2023llm}.
Specifically, current edge devices offer only a limited amount of memory capacity, ranging from a few tens of megabytes to a few gigabytes. For example, NVIDIA Jetson Nano has 4 GB of memory and Raspberry Pi 4 Model B has up to 8 GB of memory.
In contrast, large models' parameters have experienced exponential growth, reaching sizes in the hundreds of billions.
For instance, the GPT-3~\cite{kalyan2023survey} model has 175 billion trainable parameters, while the recently developed GPT-4~\cite{sanderson2023gpt} model exceeds the trillion parameter mark.
Consequently, the memory usage of these large models can easily reach tens to hundreds of gigabytes, far surpassing the memory capacity of typical edge devices.

Existing works to address the memory challenges of large model inference on edge devices can be classified into two categories. 
The first attempts to optimize the model structure to reduce the computational load through techniques including model pruning~\cite{ma2023llm,he2023structured,mostafa2019parameter}, model compression~\cite{deng2020model,liu2022multi}, model quantization~\cite{yao2021hawq} and adaptive inference~\cite{tambe2021edgebert,jin2020adabits}. 
Although these approaches significantly diminish the number of required computational operations, they often result in reduced model accuracy. Moreover, these approaches are generally tailored for specific models, thus limiting their applicability on a broader scale. 
The second optimizes the memory usage during model inference by model swapping between different storage media~\cite{huang2020swapadvisor,hao2023reaching,jiang2024slob}. 
This method initially divides the model into separate shards and selectively preloads certain shards from disk into a buffer or the memory as needed for the inference process. 
Even though memory swapping methods can reduce memory overhead, they can inadvertently prolong inference latency due to the increased frequency of I/O operations between varying storage media.

\begin{figure}[!t]
    \centering
    \begin{subfigure}{0.48\textwidth}
        \includegraphics[width=\textwidth]{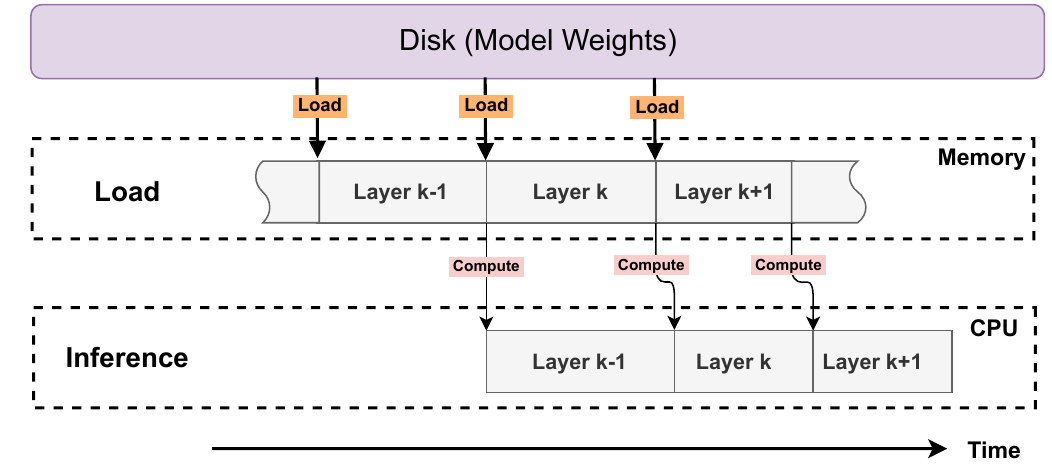}
        \caption{Standard pipeline.}
        \label{fig:first}
    \end{subfigure}
    \hfill
    \begin{subfigure}{0.48\textwidth}
        \includegraphics[width=\textwidth]{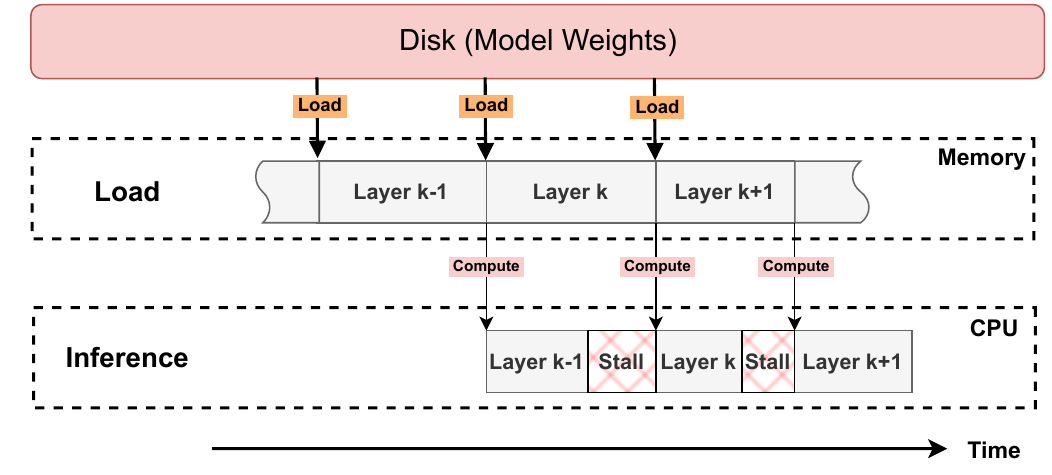}
        \caption{Pipeline stalls.}
        \label{fig:second}
    \end{subfigure}
    \caption{Standard pipeline design and pipeline stall problem.}
    \label{fig:pp4}
\end{figure}

In this paper, we envisage pipelining the model loading and inference process, hiding the latency of the loading by overlapping it with the inference. Given the ubiquity of CPUs in edge devices, the loading process involves loading model weights from disk to memory, and the inference process refers to performing inference on CPUs. Fig. \ref{fig:first} illustrates the standard pipeline design, which loads model weights from disk to memory at a layer granularity. A transformer layer consists of the multi-head self-attention and the position-wise feed-forward network. Given that the transformer model is comprised of sequential layers, it performs the loading and inference process layer-by-layer. Inference in memory begins immediately following the loading of the initial layer from disk, with the subsequent layer being loaded concurrently. The standard pipeline formed by this process allows inference to commence prior to the complete model being loaded, thus reducing the latency.

Although we are not the first to apply pipeline to large model inference on edge devices, we attempt to solve two challenges that have not been addressed by existing works, such as PipeEdge~\cite{hu2022pipeedge} and PipeSwitch~\cite{bai2020pipeswitch}. The first challenge is that pipeline schemes do not reduce the memory requirements of inference. For instance, PipeEdge employs pipeline parallelism, leveraging under-utilised or idle distributed edge resources to enhance inference performance across diverse edge devices. Although this approach enhances inference speed, it lacks memory optimization and does not reduce the memory requirements for inference. The second challenge is that the deployment of pipeline on edge may lead to serious pipeline stalls, because of the huge gap between the model loading and execution latency. Our experiments in \S\ref{sec:back_char} demonstrate that the loading latency is generally an order of magnitude larger than the inference latency, leading to the pipeline stall issue illustrated in Fig. \ref{fig:second}.

To resolve two issues mentioned above, we develop \mechanism, a memory-efficient pipeline execution mechanism to streamline the loading and inference process with per-layer granularity. This mechanism incorporates dynamic management of memory with timely destruction of model weights that have been inferred, significantly reducing the memory usage of model inference. And by engaging parallel model loading, \mechanism overlaps multiple inference time within a single loading interval to minimise the pipeline stalls, consequently decreasing inference latency. 

Building upon this mechanism, we introduce \name, an innovative framework for large model inference on edge devices. This framework integrates three main components. Firstly, the \LP evaluates the performance and memory utilization of each individual model layer for a given transformer model. Next, the \PP utilizes the profiling data from \LP to devise a \mechanism execution schedule within different memory constraints. Lastly, the \EE determines the execution strategy from the schedule based on the current memory constraints of the edge device and executes \mechanism inference.


We conduct experiments with four transformer-based models, including BERT-Large, GPT-2-Base, ViT-Large and GPT-J, on our CPU cluster server. These four models vary in size from a few hundred megabytes to a dozen gigabytes. Through a comprehensive performance and memory footprint evaluation, \name achieves up to \(4.24\times\) increase in inference speed and \(86.7\%\) lower memory footprint than PipeSwitch for BERT and ViT models, \(2.58\times\) increase in inference speed and \(90.3\%\) lower memory footprint for GPT-style models. We also evaluate \name under different memory constraints and it works well in all test environments with all results meeting service level objective (SLO) expectations.


Our contributions are highlighted as follows.
\begin{itemize}
    \item We propose \mechanism, a memory-efficient pipeline mechanism designed to reduce the memory footprint and latency during model inference.
    \item We present \name, a framework based on the \mechanism to mitigate memory usage and pipeline stall for large model inference on edge devices.
    \item We implement a system prototype of \name and evaluate it on several transformer models. Experiments show that our method achieves a \(4.24\times\) speedup while reducing memory footprints by \(90.3\%\).
\end{itemize}

\begin{figure*}[!t]
    \centering
    \begin{minipage}{0.48\textwidth}
        \centering
        \includegraphics[width=\linewidth]{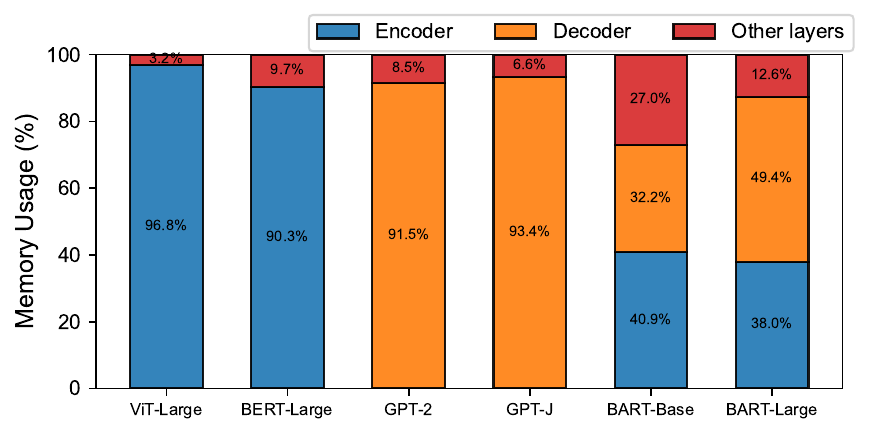}
        \caption{Decomposition of layers' memory usage.}
        \label{fig:mem}
    \end{minipage}
    \hfill
    \begin{minipage}{0.48\textwidth}
        \centering
        \includegraphics[width=\linewidth]{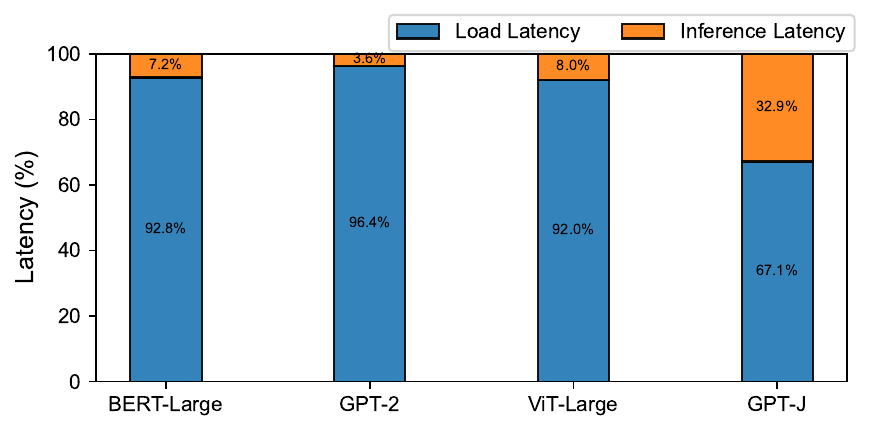}
        \caption{Decomposition of loading and inference latency.}
        \label{fig:time}
    \end{minipage}
\end{figure*}
\section{Background and Motivation} \label{sec:back}

\subsection{Transformer Model Structure}
The architectural makeup of transformer models is the basis for developing pipeline inference strategies on edge devices. A typical transformer model comprises embedding, encoder, decoder, pooling, and additional specialized layers. Based on their architectural configurations, they can be classified into three primary categories: encoder-decoder, encoder-only and decoder-only models. Encoder-decoder models, such as BART~\cite{lewis2019bart} and T5~\cite{raffel2020exploring} integrate both encoder and decoder layers. BART is architected for complex sequence-to-sequence tasks, and T5 generalizes this capability with a comprehensive text-to-text methodology applicable to a wide array of NLP challenges. Encoder-only models like BERT~\cite{devlin2018bert} employ a series of encoder layers to interpret input data, ideally suited for tasks that do not involve generating sequences. Meanwhile, Vision Transformer (ViT) presents an innovative adaptation of encoder-only architecture to analyze sequences of image patches, eschewing the conventional decoder layout found in text-centric models. Conversely, decoder-only models such as GPT rely exclusively on decoder layers, focusing on the generation of new content by drawing on recognizable patterns. 

\subsection{Characteristics of Transformer-based Models} \label{sec:back_char}
In order to design an efficient pipeline scheme, we conduct experiments to characterize two key aspects when the model performs forward computation: the allocation of memory among transformer model layers and the latency of model loading and model inference. 
We first evaluate the memory allocation in five kinds of transformer models, including ViT-Large, BERT-Large, GPT-2, GPT-J and BART (BART-Base and BART-Large), which cover all three categories of Transformer models. Additionally, we evaluate the time requirements of loading and inference for various transformer models, including BERT-Large, GPT-2, ViT-Large and GPT-J, by performing standard model inference. All the experiments are conducted on Intel(R) 193 Xeon(R) Gold 6248R CPU.

\noindent\textit{\textbf{TestCase1: Memory distribution.}} To understand the allocation of memory across layers, we conduct experiments with five kinds of transformer models. Typically, transformer-based models are characterized by their extensive reliance on attention mechanisms, necessitating substantial memory to accommodate attention scores and intermediate representations, particularly within encoder or decoder layers. Fig. \ref{fig:mem} delineates the memory usage distribution across different layers for five kinds of prevalent transformer variants, revealing that encoder or decoder layers predominate, consuming between 70\% to 95\% of the total memory. Notably, the memory consumption attributed to these layers escalates with the model's overall size. For instance, BART-Large necessitates approximately 14.4\% more memory relative to BART-Base.

\vspace{6pt}
\begin{thmbox}[M]{\textbf{Observation~\uppercase\expandafter{\romannumeral1}}}
    \emph{For general transformer-based models, the encoder or decoder layers occupy the largest memory footprint.}
\end{thmbox}
\vspace{4pt}
\noindent\textit{\textbf{TestCase2: Latency evaluation.}} To evaluate the latency of model loading and inference, we run standard model inference processes for four transformer models on CPU. Generally, transformer models exhibit considerably higher latency during layer loading compared to layer inference. Through our experiments, as depicted in Fig. \ref{fig:time}, we observe that, for the first three smaller models (each with a memory footprint around 1 GB), the layer loading period substantially exceeds the inference time, by roughly an order of magnitude. Conversely, for the larger GPT-J model (12 GB), the layer loading duration is approximately twice that of the inference time. Subsequently, such disparities contribute to a significant portion of the computational process, between 60\% to 80\%, being spent idle during typical pipeline execution, underlining a serious pipeline stall issue, as shown in Fig. \ref{fig:second}.

\vspace{6pt}
\begin{thmbox}[M]{\textbf{Observation~\uppercase\expandafter{\romannumeral2}}}
    \emph{For general transformer-based models, loading latency is much larger than the inference latency, resulting in the execution process being stalled during most of inference time.}
\end{thmbox}
\vspace{4pt}

\subsection{Implications}
Our experiments in \S\ref{sec:back_char} analyze the time distribution and memory usage of transformer-based large models during model inference. 
{\textbf{Observation~\uppercase\expandafter{\romannumeral1}}} suggests that a targeted focus on either the encoder or decoder layers is pivotal for optimizing memory management in our pipeline infrastructure. {\textbf{Observation~\uppercase\expandafter{\romannumeral2}}} underscores the necessity of adopting a parallel loading strategy by overlapping multiple inference times with a single loading time within our pipeline scheme, to efficiently mitigate pipeline stalls. In summary, we progress our design by addressing the following challenges: (1) memory challenge on edge devices; (2) pipeline stall problem caused by the huge gap between loading and inference latency.

\begin{figure}[!t]
    \centering
    \includegraphics[width=1.0\linewidth]{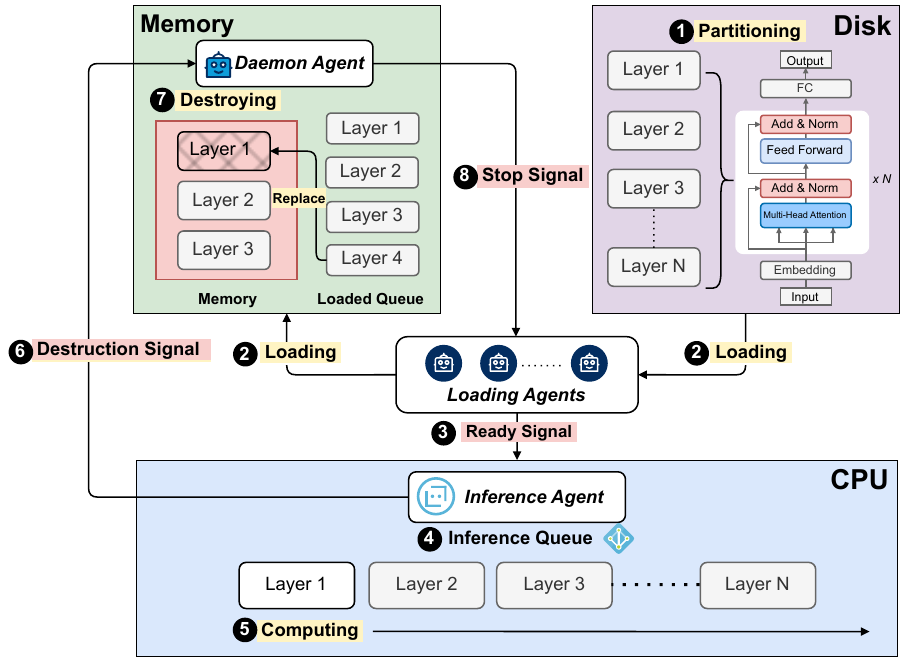}
    \caption{The overview and workflow of \mechanism.}
    \label{overview}
\end{figure}

\begin{figure}[!t]
    \centering
    \includegraphics[width=.99\linewidth]{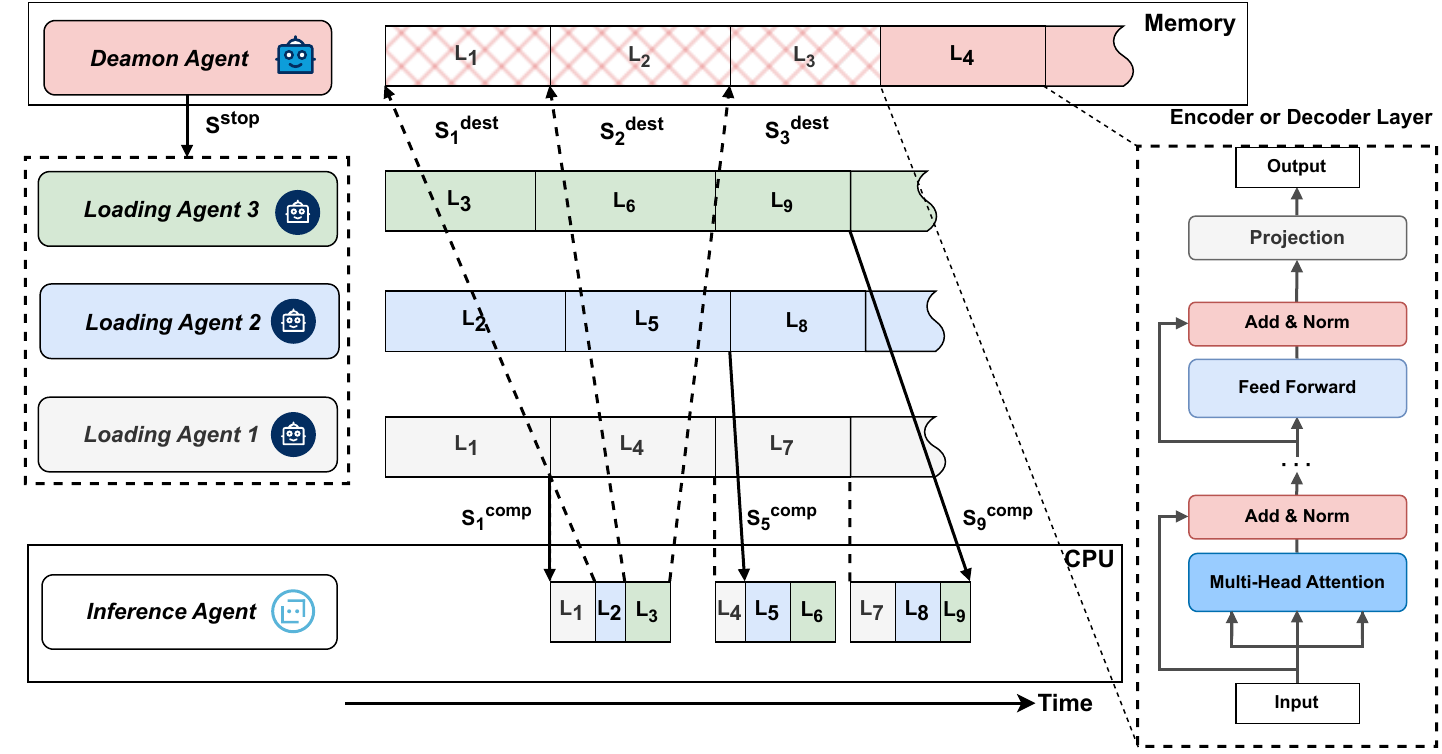}
    \captionsetup{justification=centering}
    \caption{\mechanism with three \LAs.}
    \label{fig:pp2}
\end{figure}


\section{PipeLoad: A Memory-Efficient Pipeline Execution Mechanism}

\subsection{Overview}

We present \mechanism, a memory-efficient pipeline execution mechanism to reduce memory footprint and latency during model inference on edge devices. There are three core workers in \mechanism mechanism: multiple \LAs, one \IA and one \DA. \LAs work in parallel to load model layers from disk to memory, reducing inference latency. The \IA simultaneously executes computations on these loaded layers sequentially in CPU, guaranteeing the model's predictive accuracy and minimizing pipeline stalls. The \DA maintains a queue of loaded layers in memory, detects memory usage and destroys memory space for specific layers at a specific point to reduce memory overhead. Three workers communicate with each other through a signalling mechanism that facilitates the realization of whole memory-efficient pipeline.

Fig. \ref{overview} illustrates the overall workflow of \mechanism, including the loading and inference process of model layers as well as the signaling mechanism. Before performing pipeline inference, we \circled{1} adopt a layer-based model partitioning scheme to pre-process the model weights. Multiple \LAs constantly \circled{2} load specific layers from disk into memory in parallel. Once a model layer is successfully loaded, the corresponding \LA \circled{3} transmits a computation ready signal that corresponds to this layer to \IA, indicating that this layer is ready for computation. \IA \circled{4} maintains an inference queue in CPU that decides which layer will be processed next, ensuring that model inference respects the original sequence of layers. Upon receiving the computation ready signal, \IA \circled{5} performs forward computation only if all preceding layers have been computed. Following computation of the layer, \IA \circled{6} issues a memory destruction signal to \DA, notifying it to destroy the memory space of the layer. \DA then \circled{7} destroys the memory space occupied by the layer to reserve enough space for other layers. When memory usage is about to exceed or has exceeded the memory constraints of the edge device, \DA \circled{8} sends a stop signal to all \LAs, pausing their loading operations until sufficient memory space is available.

\begin{figure*}
    \centering
    \includegraphics[width=.99\linewidth]{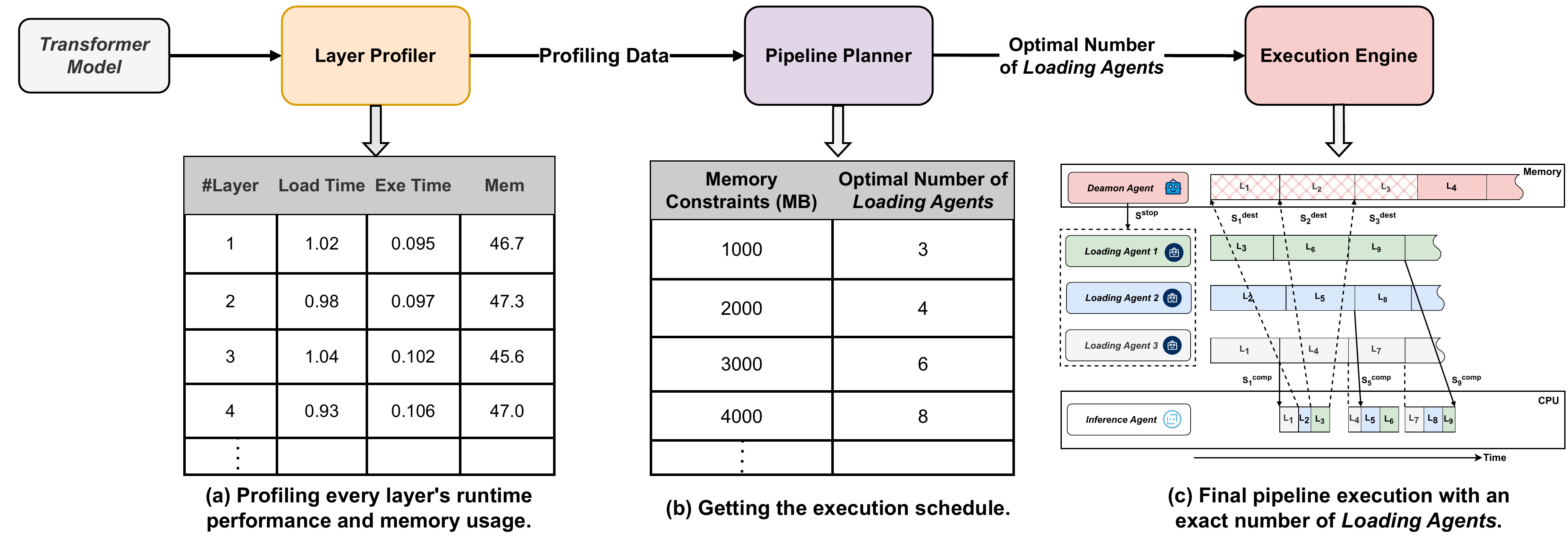}
    \captionsetup{justification=centering}
    \caption{\name system architecture.}
    \label{fig:pp1}
\end{figure*}
\subsection{Case Study}
Fig. \ref{fig:pp2} presents a simple case of \mechanism with three \LAs. Based on the characteristics of transformer model layers, we adopt a layer-based model partitioning scheme. In our scheme, we methodically segment the general transformer model architecture into its constituent layers: embedding layers, encoder layers, decoder layers and other layers. Among these layers, we focus only on the encoder and decoder layers that occupy most of the model weights in \mechanism mechanism design. 

For simplicity, we show only three computation ready signals and three memory destruction signals in Fig. \ref{fig:pp2}. And for the sake of clarity, three \LAs are symbolized as \(LA_1, LA_2, LA_3\). Model layers are signified as \( L_k \) where \( k \) represents the index within the total number of layers, denoted by \( N \). Symbols \( S^\text{comp}_k \), \( S^\text{dest}_k \) and \( S^\text{stop} \) respectively represent computation ready signal for layer \( L_k \), memory destruction signal for layer \( L_k \) and loading stop signal. During the implementation of \mechanism, the $i$-th \LA is assigned a subset of model layers, following the distribution \( L_{i + jm} \), where \( i \) ranges from 1 to \( m \), with \( m \) representing the total number of \LAs, and \( j \) represents an iterative index, ranging from 0 to $\lfloor (N-i)/m \rfloor$ ($i+jm \leq N$ and $j \in \mathbb{N} $). In this case, \( LA_1 \) is responsible for layers $(L_1, L_4, L_7, \ldots)$, \( LA_2 \) for layers $(L_2, L_5, L_8, \ldots)$ and \( LA_3\) for layers $(L_3, L_6, L_9, \ldots)$. This layer allocation method is designed to minimize pipeline stalls since we can overlap the inference time of three layers with the loading time of a single layer. 

As shown in Fig. \ref{fig:pp2}, the three \LAs commence the parallel loading process. As the layer \( L_1 \) is fully loaded to memory, \( LA_1 \) issues the computation ready signal, \( S^\text{comp}_1 \) to \IA. After receiving $S^\text{comp}_1$, \IA starts to perform forward computation for $L_1$. If \IA receives $S^\text{comp}_2$ or $S^\text{comp}_3$ first, the inference queue in CPU will ensure that the layers are computed in the correct order. Simultaneously, after \( LA_2 \) and \( LA_3 \) load $L_2$ and $L_3$ as well as sending computation ready signals to \IA, the three \LAs are able to continue loading the respective next layers $L_4, L_5, L_6$ into memory. Following the computation on \( L_1 \), \IA issues the memory destruction signal, \( S^\text{dest}_1 \) to \DA, which initiates the process of de-allocating memory space for $L_1$. The loading and inference process for the subsequent layers is similar. In addition, when \DA detects the system memory usage has exceeded the memory constraints of the edge device, it sends the loading stop signal, \( S^\text{stop} \) to all \LAs, pausing their layers' loading operation until sufficient memory space is available.

\section{\name: A framework to Optimize Large Model Inference on Edge Devices}


Fig. \ref{fig:pp1} presents \name system architecture, a comprehensive framework designed to enhance the performance and reduce memory usage of large model inference in edge computing environments. Specific modules within \name comprise \LP, \PP, and \EE. This framework encapsulates methodologies for evaluating layer efficiency, deploying an optimal execution schedule, executing the memory-efficient pipeline, \mechanism and aims to collaborate diverse elements essential for optimizing model execution in resource-constrained settings, such as memory usage, latency and execution strategy. 

\subsubsection{\LP}

Fig. \ref{fig:pp1}a presents some possible results of \LP, which serves as the foundation of our system architecture. The \LP's primary function is to profile each layer within a given transformer model to gauge runtime performance and memory usage. Through a pre-run of standard model inference, this profiling enables the accurate measurement of loading time, computation time and memory size for every individual layer of the given model.

\subsubsection{\PP}

Utilizing the data generated by the \LP, the \PP develops a \mechanism execution schedule that includes several optimal execution strategies under different memory constraints, as shown in Fig. \ref{fig:pp1}b. Firstly, drawing from the profiling insights encompassing layer's memory footprint along with layer's load and compute duration for the given model, the planner determines a reasonable range for the number of \LAs in conjunction with different memory constraints. In general, more \LAs means fewer pipeline stages, \textit{i.e.}, less latency, but more encoder or decoder layers are reserved in memory, \textit{i.e.}, more memory overhead. Next, the planner pre-runs the \mechanism within the range of the number of \LAs to obtain the exact number of \LAs under different memory constraints and finally outputs the execution schedule.


\subsubsection{\EE}

Finally, upon establishing the execution schedule, the inference of \mechanism with an exact number of \LAs will be executed in the \EE based on the current memory constraints of edge device, as shown in Fig. \ref{fig:pp1}c. This includes actual pipeline inference execution facilitated by the specific number of \LAs, one \IA, one \DA and signalling mechanism.

\section{Evaluation}


\begin{table*}[!t]
\centering
\caption{Model Configurations.}
\label{tab:model_configs}
\renewcommand\arraystretch{1.5}
\resizebox{0.93\linewidth}{!}{
\begin{tabular}{|c |c |c |c |c |c |c|}
\hline
\textbf{Model}& \textbf{Parameters Size (Millions)} &\textbf{Types of Layers }&\textbf{Number of Layers} &\textbf{Data Type} &\textbf{Memory (Layers/Total) (MB)} &\textbf{Memory per Layer (MB)}\\ \hline
\hline
ViT-Large &304 &encoder &24 &FP16 &582 / 601 &25\\
\hline
GPT-2-Base &355 &decoder &24 &FP32 &1223 / 1433 &51\\
\hline
BERT-Large &340 &encoder &24 &FP32 &1317 / 1627 &55\\
\hline
GPT-J &6000 &decoder &28 &FP32 &11535 / 12354 &412\\
\hline
\end{tabular}
}
\end{table*}

\begin{table*}[!t]
\centering
\caption{Performance comparison.}
\label{tab:model_speedup}
\renewcommand\arraystretch{1.5}
\resizebox{0.93\linewidth}{!}{
\begin{tabular}{|c|c|c|c|c|c|c|c|c|c|}
\hline
\multirow{2}*{\textbf{Model}}&\multicolumn{1}{|c|}{\textbf{Baseline}}&\multicolumn{2}{|c|}{\textbf{PipeSwitch}}&\multicolumn{2}{|c|}{\textbf{\mechanism with 2 \textsl{LAs}}}&\multicolumn{2}{|c|}{\textbf{\mechanism with 4 \textsl{LAs}}}&\multicolumn{2}{|c|}{\textbf{\mechanism with 6 \textsl{LAs}}}\\ 
\cline{2-10}
    &Latency (ms)&Latency (ms)&Speedup&Latency (ms)&Speedup&Latency (ms)&Speedup&Latency (ms)&Speedup\\ 
\hline\
BERT-Large&15891.5&14897.1&1.067&7720.8&2.058&4621.8&3.438&3510.7&\textbf{4.527}\\ 
\hline
GPT-2-Base&1659.5&2457.9&0.675&1704.7&0.974&1396.1&1.189&1121.4&\textbf{1.480}\\ 
\hline
ViT-Large&345.0&157.3&2.193&90.8&3.799&56.8&6.070&43.2&\textbf{7.978}\\ 
\hline
GPT-J&31330.9&76494.6&0.410&51003.3&0.614&33487.2&0.936&29640.9&\textbf{1.057}\\
\hline
\end{tabular}
}
\end{table*}

\begin{table*}[!t]
\centering
\caption{Memory footprints comparison.}
\label{tab:model_memory}
\renewcommand\arraystretch{1.6}
\resizebox{0.93\linewidth}{!}{
\begin{tabular}{|c|c|c|c|c|c|c|c|c|c|}
\hline
\multirow{2}*{\textbf{Model}}&\multicolumn{1}{|c|}{\textbf{Baseline}}&\multicolumn{2}{|c|}{\textbf{PipeSwitch}}&\multicolumn{2}{|c|}{\textbf{\mechanism with 2 \textsl{LAs}}}&\multicolumn{2}{|c|}{\textbf{\mechanism with 4 \textsl{LAs}}}&\multicolumn{2}{|c|}{\textbf{\mechanism with 6 \textsl{LAs}}}\\
\cline{2-10}
    &Memory footprint (MB)&Memory footprint (MB)&Ratio&Memory footprint (MB)&Ratio&Memory footprint (MB)&Ratio&Memory footprint (MB)&Ratio\\
\hline
BERT-Large&1627.3&1689.2&1.038&457.1&\textbf{0.281}&661.5&0.407&930.8&0.572\\
\hline
GPT-2-Base&1433.8&1436.8&1.002&387.5&\textbf{0.270}&518.8&0.362&649.9&0.453\\
\hline
ViT-Large&600.9&626.6&1.043&60.8&\textbf{0.101}&110.2&0.183&159.4&0.265\\
\hline
GPT-J&12354.0&12468.6&1.009&1668.6&\textbf{0.135}&2455.4&0.199&3242.2&0.262\\
\hline
\end{tabular}
}
\end{table*}

\subsection{Experimental Setup}

\subsubsection{Workloads}
For estimating the memory-efficient pipeline execution mechanism, \mechanism, we focus on a quartet of transformer models:
\begin{inparaenum}[\itshape i\upshape)] 
    \item one NLP model: BERT-Large;
    \item one CV model: ViT-Large;
    \item and two generative text language models: GPT-2-Base and GPT-J.
\end{inparaenum}
These four transformer models have different sizes, from a few hundred megabytes to a dozen gigabytes. Each of their configurations is shown in TABLE \ref{tab:model_configs}, where the number of layers is the number of encoder or decoder layers, excluding other layers such as embedding layers and pooling layers, memory (layers / total) indicates that the memory footprint of encoder or decoder layers accounts for the total memory of the model and memory per layer represents the average memory footprint per encoder or decoder layer. 

\subsubsection{Baselines}
In performance and memory usage evaluation, we focus on evaluating four transformer models mentioned above in \EE. And the engine provides three distinct operational modes: Baseline (non-pipeline), PipeSwitch, and our designed \mechanism with optional \LAs. In particular, the workflow of Baseline is the normal process of loading model first and then inferring it, and the workflow of PipeSwitch is basically the same as the standard pipeline.

\subsubsection{Metrics}
We use two performance metrics, latency and memory footprints. 
In the context of BERT and ViT models, latency is defined as the end-to-end time taken to generate an output with single inference; for GPT-style models, latency defines as the end-to-end output generation time for a given input prompts and a given output tokens length.
Memory footprints is quantified as the maximum memory occupation by the model throughout its execution lifecycle.

\subsubsection{Testbeds}
We conduct our experiments on a server consisting of Intel(R) Xeon(R) Gold 6248R CPUs. We deploy a controlled and consistent environment with \texttt{docker} that imposes limits on resource usage, including limiting the number of CPU cores to a maximum of 8 and restricting memory size through \texttt{docker --memory} command, to simulate resource-constrained scenarios on edge devices.

\subsection{Evaluation of Performance and Memory Footprints}
We evaluate the performance and memory footprints of \mechanism with 2, 4 and 6 \LAs and compare them to baseline and to PipeSwitch. We choose these three numbers of \LAs since they are essentially factors of the number of encoder or decoder layers in four transformer models. The performance and memory footprints test results are shown in TABLE \ref{tab:model_speedup} and TABLE \ref{tab:model_memory} respectively, where \textsl{LAs} is an acronym for \LAs. In order to show the optimisation results more directly, we add two metrics in tables respectively, the speedup and the ratio, with their expressions are as follows:

\[ \text{Speedup} = \frac{T_{\text{baseline}}}{T_{\text{others}}} \]
\[ \text{Ratio} = \frac{M_{\text{others}}}{M_{\text{baseline}}} \]


\noindent where $T_{\text{baseline}}$ and $T_{\text{others}}$ represent the latency of baseline and latency of other methods and $M_{\text{others}}$ and $M_{\text{baseline}}$ indicate the memory consumption of other methods and memory consumption of baseline. 

\begin{figure*}[!t]
    \begin{subfigure}[c]{0.24\textwidth}
        \centering
        \includegraphics[width=0.99\linewidth]{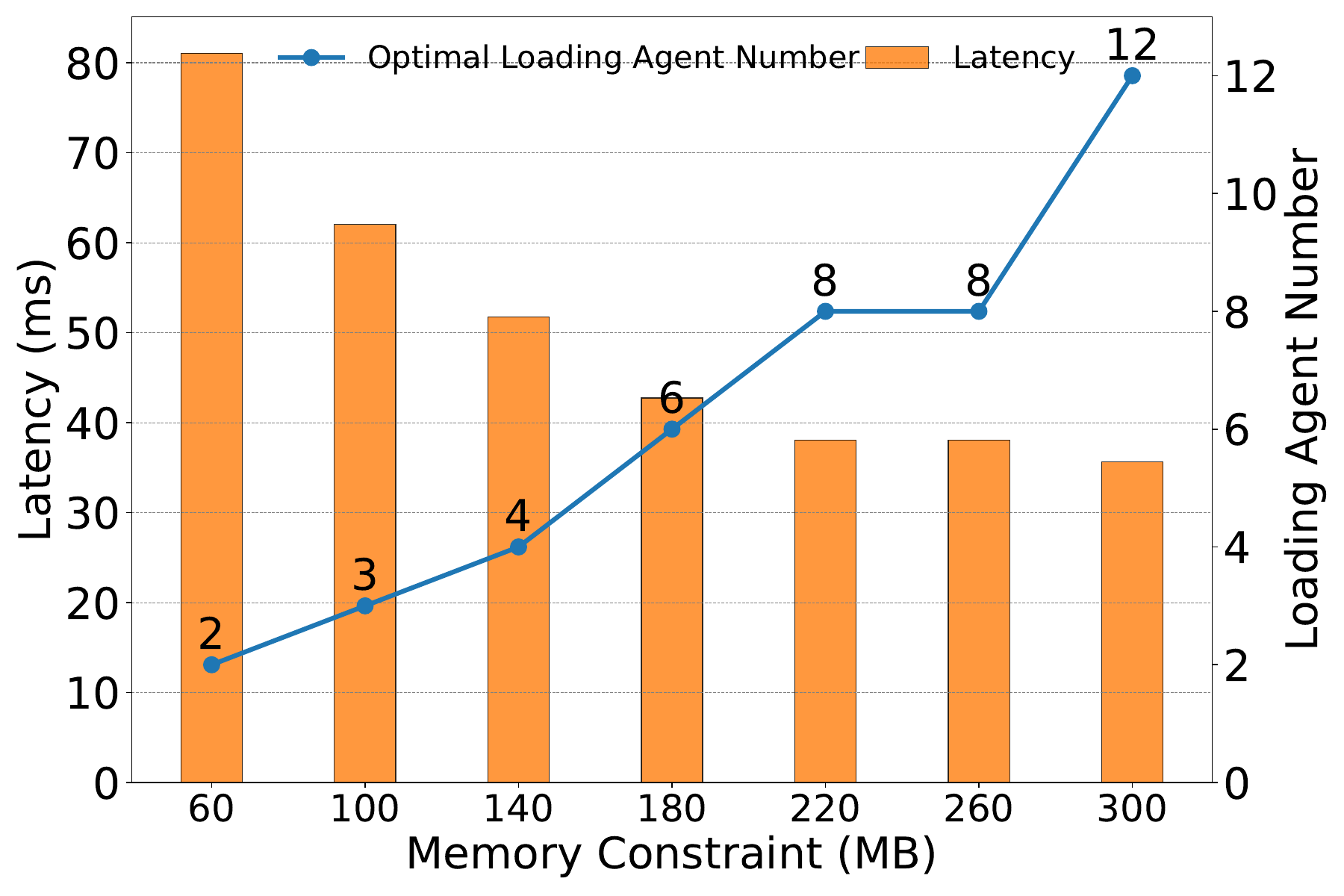}
        \caption{ViT-Large model.}
        \label{fig:vit}
    \end{subfigure}
    \hfill
    \begin{subfigure}[c]{0.24\textwidth}
        \includegraphics[width=0.99\linewidth]{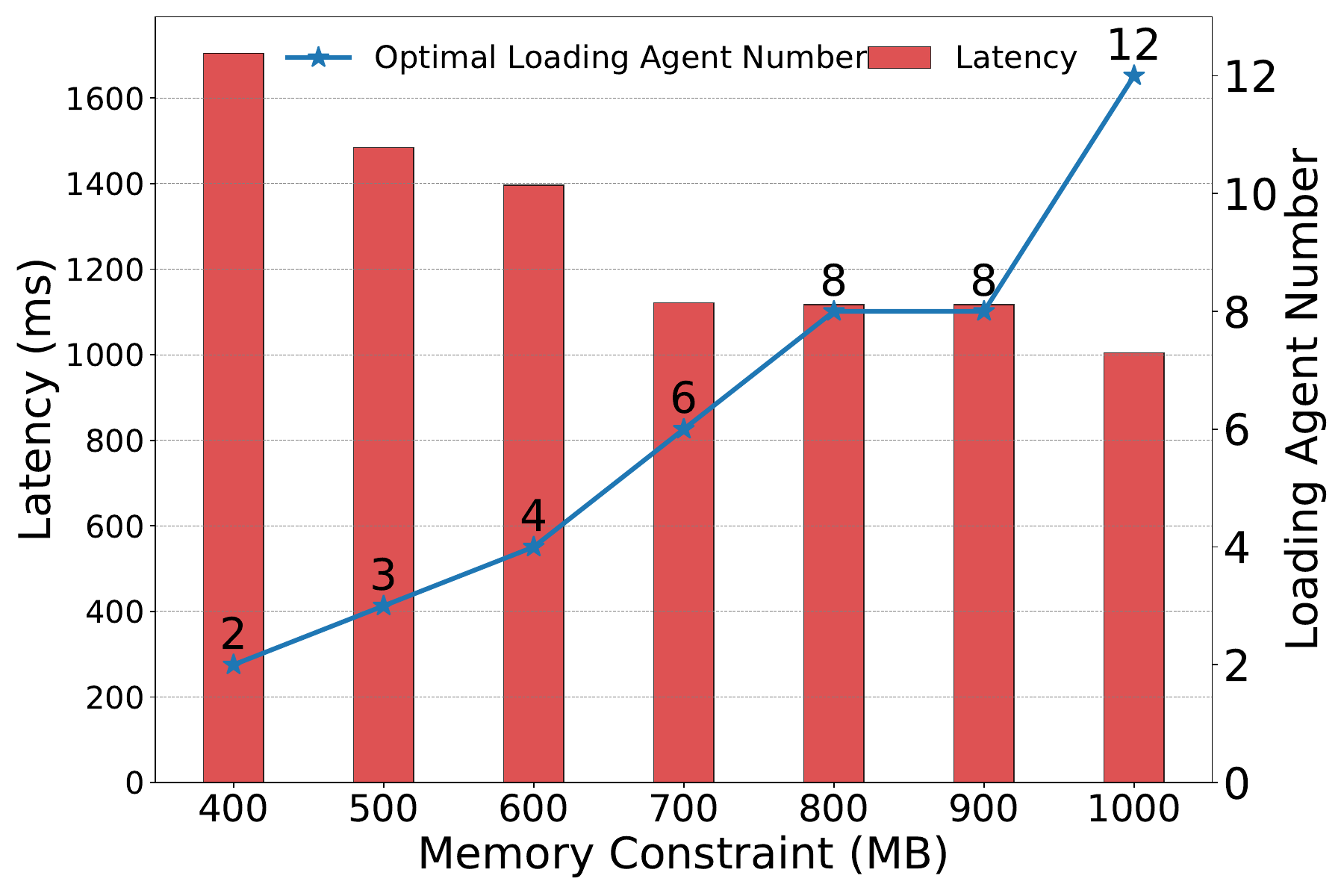}
        \caption{GPT-2-Base model.}
        \label{fig:gpt2}
    \end{subfigure}
    \hfill
     \begin{subfigure}[c]{0.24\textwidth}
        \centering
        \includegraphics[width=0.99\linewidth]{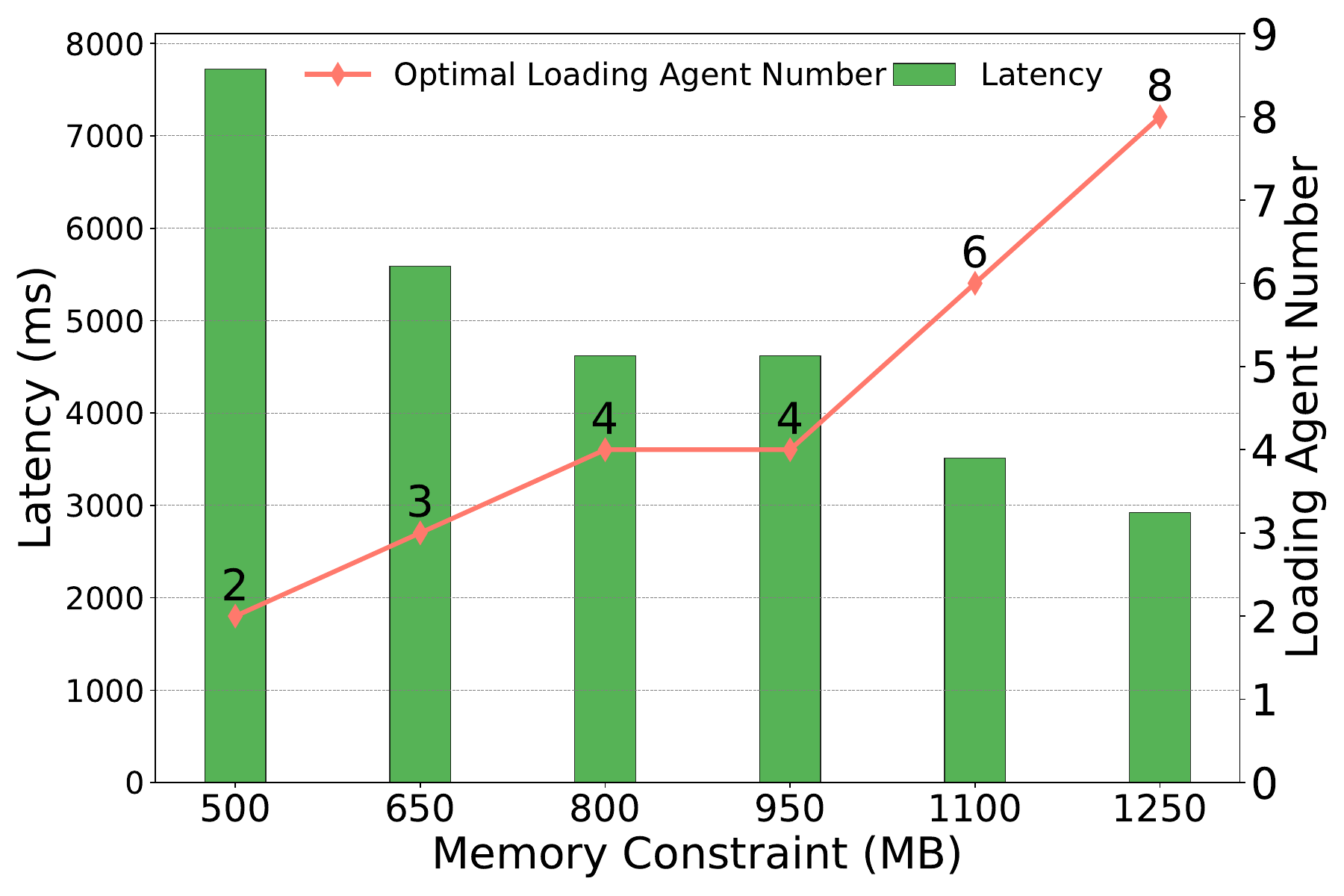}
        \caption{BERT-Large model.}
        \label{fig:bertlarge}
    \end{subfigure}
    \hfill
    \begin{subfigure}[c]{0.24\textwidth}
        \includegraphics[width=0.99\linewidth]{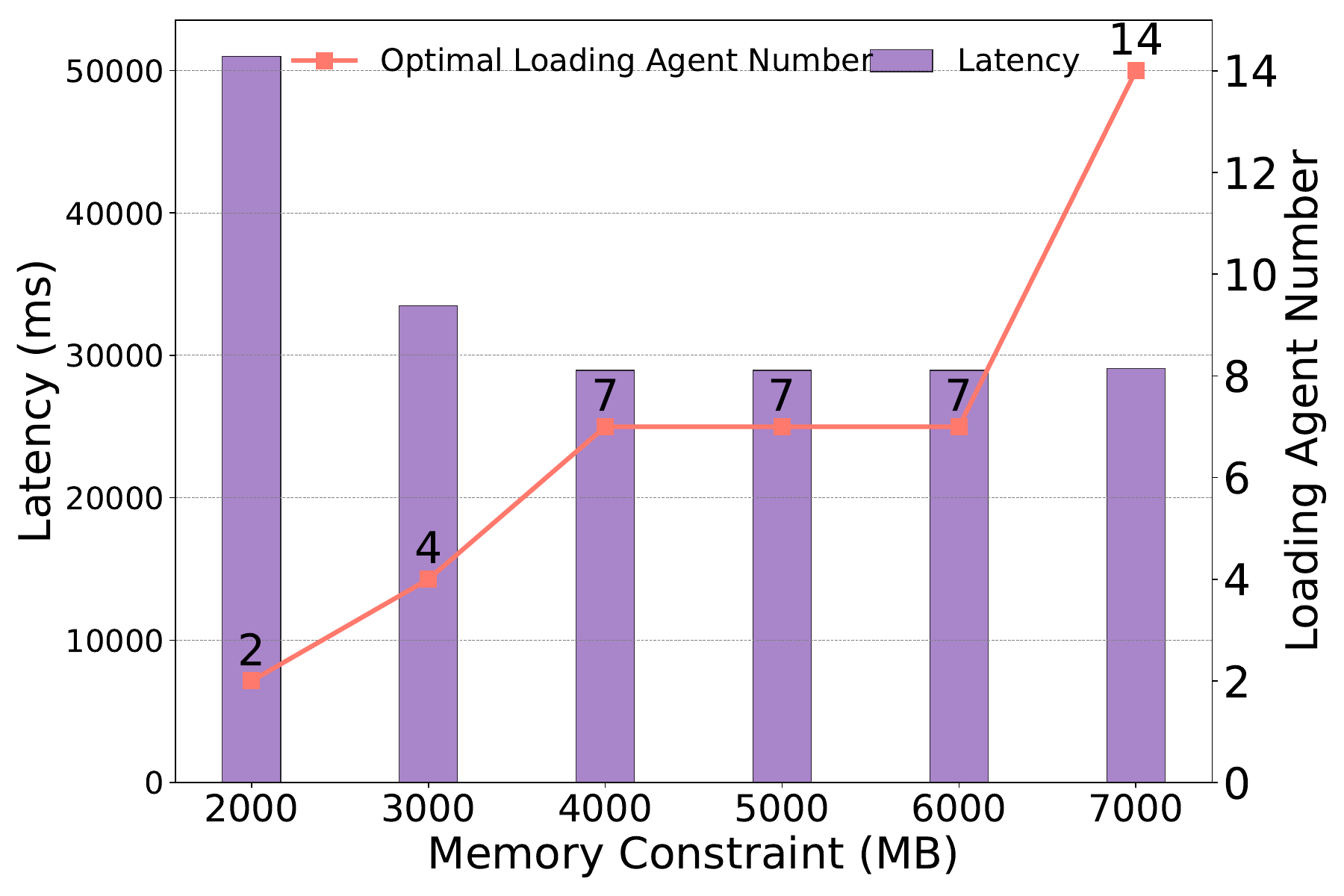}
        \caption{GPT-J model.}
        \label{fig:gptj}
    \end{subfigure}
    \caption{Models evaluation under different memory constraints.}
    \vspace{-0.15in}
\end{figure*}

\subsubsection{BERT and ViT Models Analysis}
For BERT and ViT models, we evaluate them with a single inference since they can generate outputs through loading and inference in a single pass.
According to TABLE \ref{tab:model_speedup} and TABLE \ref{tab:model_memory}, \mechanism with multiple \LAs indicates a promising trend of decreasing memory usage and latency compared to the PipeSwitch implementation. For BERT-Large, the speedup improvement is \(1.93 \sim 4.24\times\) and the memory footprint reduction is \(44.9\% \sim 73.0\%\). For ViT-Large, the speedup improvement is \(1.73 \sim 3.64\times\) and the memory footprint reduction is \(74.0\% \sim 86.7\%\). The smaller proportion of memory footprint reduction for BERT model compared to the ViT model is mainly due to the fact that the embedding and pooling layers of BERT-Large have a much larger portion, about \(20\% \) while ViT-Large about \(1.5\% \). As we increment the number of \LAs, the speedup is also significantly increasing while the degree of memory footprint reduction is slowly decreasing. This result is also as expected, since more \LAs means less pipeline stages and more encoder or decoder layers saved in memory. Specifically, adding one \LA implies one additional layer saved in memory.

\subsubsection{GPT-2 and GPT-J Models Analysis}
For the two GPT-style transformer models, we evaluate them for a given input prompts (number of tokens = 4) and for a given output tokens (length of tokens = 8). 
As shown in TABLE \ref{tab:model_speedup} and TABLE \ref{tab:model_memory}, \mechanism with multiple \LAs produces excellent results in reducing memory usage and good results in decreasing latency compared to the PipeSwitch. For GPT-2-Base, the speedup improvement is \(1.44 \sim 2.20\times\) and the memory footprint reduction is \(13.7\% \sim 72.9\%\). For GPT-J, the speedup improvement is \(1.50 \sim 2.58\times\) and the memory footprint reduction is \(74.6\% \sim 90.3\%\). \mechanism works better in GPT-J slightly better because its proportion of decoder layers' size to the total model size is higher. And for the same reasons as the previous two models, with the number of \LAs growing, the speedup is increasing while the degree of memory footprint reduction is slowly decreasing. However, their improvements in execution speed are less effective in comparison to the baseline when the number of \LAs is low ($ \le 4 $). This is because the GPT-style transformer model loads memory only once for non-pipeline execution but performs inference multiple times (one inference for each token) while \mechanism and other pipeline methods require one loading and inference operation for each token, thus increasing latency when there are a large quantity of tokens.  

\subsection{Evaluation under different Memory Constraints}
We evaluate the performance of \name under different memory constraints. In addition, we measure the latency and the corresponding optimal number of \LAs.

\subsubsection{ViT and BERT Models Analysis}
Fig. \ref{fig:vit} and Fig. \ref{fig:bertlarge} show the evolution of latency and optimal number of \LAs with respect to memory constraints for ViT-Large model and BERT-Large model. Across the experiments, a trend is the gradual increase in the optimal number of \LAs, the decrease in the latency in correlation with the augmentation of memory limits. Specifically, the latency dropped from 81 ms at the 60 MB memory limit to 36 ms at 300 MB memory limit for ViT-Large, a reduction of \(55.6\% \) and from 7721 ms at the 500 MB memory limit to 2923 ms at the 1250 MB memory limit for BERT-Large, a reduction of \(62.1\% \). All above results are as expected, since higher memory availability allows for more \LAs.

\subsubsection{GPT-2 and GPT-J Models Analysis}
Fig. \ref{fig:gpt2} and Fig. \ref{fig:gptj} show the evolution of latency and optimal number of \LAs with respect to memory constraints for GPT-2-Base model and GPT-J model. Overall, their trends of latency and optimal number of \LAs are the same as for the previous two models, from 1705 ms at the 400 MB memory limit to 1004 ms at 1000 MB memory limit for GPT-2-Base, a reduction of \(41.1\% \), and from 51003 ms at the 2000 MB memory limit to 29074 ms at the 7000 MB memory limit for GPT-J, a reduction of \(43.0\% \). 
\section{Related Work}

\textbf{Memory Optimization.} PQK~\cite{kim21m_interspeech} is a novel model compression method, designed expressly for edge devices with constrained computational resources. This method combines pruning, quantisation, and knowledge distillation processes to fabricate a model that is both lightweight and energy-efficient. Keivan \textit{et al.} address the memory challenges for large model inference under memory constraints by storing model parameters in flash memory and bringing them on demand to DRAM and introduce techniques including windowing and row-column bundling to optimize data transfer and memory usage. STI~\cite{guo2023sti} is a memory optimization architecture through model sharding and elastic pipeline, which employs a preload buffer to optimize resource utilization for large model inference tasks on mobile devices. Our work is complementary, focusing on minimizing inference latency by pipeline scheme while reducing memory overhead.


\textbf{Pipeline Schemes.} Prior works have attempted to apply pipeline schemes to optimize large model inference~\cite{shi2023automatic}. DeepPlan~\cite{jeong2023fast} is an optimized pipeline system that incorporates two mechanisms, direct-host-access and GPU parallel transmission to reduce the model loading latency on the GPU and improve performance. PipeSwitch is a system designed for fine-grained time-sharing of GPU resources for deep learning applications, aiming to optimize task switching overhead and achieve near 100\% GPU utilization. This system leverages the structure and computation pattern of DNN models to enable fast context switching with millisecond-scale overhead, addressing inefficiencies in shared GPU clusters where training and inference tasks are provisioned separately. These works mainly focus on reducing inference latency but do not involve memory optimization and require the use of one or even more GPUs. In this paper, we focus on both memory and latency optimization and do not require GPU usage.

\section{Conclusion}

In this paper, we present \mechanism, a memory-efficient pipeline execution mechanism to mitigate memory overhead and address the pipeline stall issue during large model inference on edge devices. This mechanism incorporates dynamic management of memory and deploys multiple \LAs to load model weights in parallel. Based on this mechanism, we introduce \name, an innovative framework to optimize large model inference performance on edge devices. By our evaluation, \name reaches \(4.24\times\) speedup and \(86.7\%\) lower memory consumption than PipeSwitch for BERT and ViT models, \(2.58\times\) speedup and \(90.3\%\) lower memory consumption for GPT-style models. For future research, we are dedicated to applying the \name to more Transformer models and exploring its generalization and pervasiveness. For text generation large models like GPT, based on their characteristics, we target to optimize \mechanism mechanism to provide better latency reduction.

\bibliographystyle{IEEEtran}
\bibliography{reference}

\end{document}